\documentclass[prb,twocolumn,showpacs,preprintnumbers,amsmath,amssymb]{revtex4}

\usepackage{graphicx}
\usepackage{dcolumn}
\usepackage{bm}

\newcommand{\be}{\begin{eqnarray}}
\newcommand{\en}{\end{eqnarray}}
\newcommand{\ben}{\begin{eqnarray}}
\newcommand{\enn}{\end{eqnarray}}
\newcommand{\beq}{\begin{eqnarray}}
\newcommand{\eeq}{\end{eqnarray}}

\newcommand{\non}{\nonumber}

\newcommand{\ds}{\displaystyle}

\newcommand{\bb}{\mathbf}

\newcommand{\sn}{{\rm{sn\,}}}
\newcommand{\cn}{{\rm{cn\,}}}
\newcommand{\dn}{{\rm{dn\,}}}

\begin{document}

\preprint{APS/123-QED}

\title{Transport magnetic currents driven by moving kink crystal in chiral helimagnets }

\author{I.~G.~Bostrem$^1$,
Jun-ichiro Kishine$^{2}$,
A.~S.~Ovchinnikov$^1$
}

\address{
$^1$Department of Physics, Ural State University, Ekaterinburg, 620083 Russia\\
$^2$Faculty of Engineering, Kyushu Institute of Technology, Kitakyushu 804-8550, Japan
}

\date{\today}

\begin{abstract}

We show that  the bulk transport magnetic current is generated by
the moving magnetic kink crystal (chiral soliton lattice) formed in
the chiral helimagnet under the static magnetic field applied perpendicular to the helical axis.
The current is  caused by the non-equilibrium transport momentum with the kink mass being determined  by the  spin fluctuations around the kink crystal state. 
An emergence of the transport magnetic currents is then a consequence of the
dynamical off-diagonal long range order along the helical axis.
We derive an explicit formula for the  inertial mass  of the kink crystal and the current in the weak field limit.
\end{abstract}

\pacs{Valid PACS appear here}
\maketitle




How to create, transport, and manipulate  spin currents is a central
problem in the multidisciplinary field of spintronics.\cite{Zutic04}
The key theoretical concepts there include the current-driven spin-transfer torque\cite{Slonczewski96} and resultant force acting on a domain wall (DW)\cite{Aharonov-Stern}
in metallic ferromagnetic/nonmagnetic multilayers,
the dissipationless spin currents in paramagnetic spin-orbit 
coupled systems,\cite{Rashba60} and
magnon transport in textured magnetic structures.\cite{Bruno05}
A fundamental query behind the issue is how to describe transport magnetic currents.\cite{Rashba05}
Conventionally, the charge current is  defined by the product of
the carrier density  and the drift velocity  
related via the continuity equation.
In the case of spin current,
the deviation of the spin projection from its equilibrium
value plays a role of a charge. Then, an emergence of the transport  magnetic currents may be expected in  non-equilibrium state
as a manifestation of the dynamical off-diagonal long range order (ODLRO).\cite{Volovik07} Historically, D\"oring first pointed out 
that the longituidal component of the slanted  magnetic moment inside the Bloch DW  emerges as a consequence of  
translational motion of the DW.\cite{Doring48} An additional magnetic energy associated with the resultant demagnetization field is interpreted as the kinetic energy of  the wall.

Recent progress of material synthesis sheds new light on this problem.
In  a series of magnets belonging  to chiral space group 
without any rotoinversion symmetry elements,
the crystallographic chirality gives rise to
the asymmetric Dzyaloshinskii interaction  
that stabilizes either 
 left-handed or right-handed 
chiral magnetic structures.\cite{Dzyaloshinskii58} 
In these chiral helimagnets, 
magnetic field applied perpendicular  to the helical axis
stabilizes a periodic array of DWs with definite spin chirality forming kink crystal or chiral 
soliton lattice.\cite{Kishine_Inoue_Yoshida2005}
In this paper,  we demonstrate that the magnetic transport analogous to D\"oring effect\cite{Doring48} occurs in the moving kink crystal of chiral helimagnets and serves an example of the dynamical ODLRO in non-equilibrium state. An essential point is that the kink crystal state has  a degeneracy originating from the translational symmetry.
Consequently, the transport momentum has a form $M\dot{X}$, where $M$ and $X$ represent 
the kink mass and the collective  coordinate of the kink  in the laboratory frame. The kink crystal behaves as a heavy object with the inertial mass $M$.

We start with a spin Hamiltonian describing the chiral helimagnet,
\be
{\cal{H}}&=&-J\sum_{<i,j>}\bb{S}_i\cdot\bb{S}_{j}+\bb{D}\cdot\sum_{<i,j>}\bb{S}_i\times\bb{S}_{j}
-\tilde{\bb{H}}\cdot\sum_{i}\bb{S}_i,\non\\\label{lattH}
\en
where the first term represents the ferromagnetic coupling with the strength $J>0$ between
the nearest neighbor $\bb{S}_i=S(\cos\theta_i,\sin\theta_i\cos\varphi_i,\sin\theta_i\sin\varphi_i)$
and $\bb{S}_j$, where $\theta_i$ and $\varphi_i$ denote the local polar coordinates.
The second term represents the parity-violating Dzyaloshinskii interaction restricted to the nearest neighbor pairs of
the adjacent ferromagnetic planes, characterized by the  the mono-axial vector $\bb{D}=D\hat{\bb{e}}_x$
along a certain crystallographic chiral axis (taken as the $x$-axis).
The third term represents the Zeeman coupling with the magnetic field $\tilde{\bb{H}}=2\mu_B H\hat{\bb{e}}_y$ applied {\it perpendicular} to the chiral axis.
When $H=0$,  the long-period incommensurate helimagnetic structure is stabilized
with the definite chirality (left-handed or right-handed) fixed by the direction of the mono-axial $\bm{D}$-vector.

\begin{figure}[h]
\includegraphics[width=85mm]{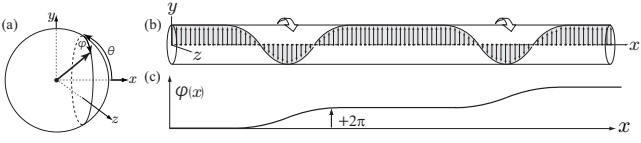}
\caption{(a) Polar coordinates in the laboratory frame.
(b) Formation of the magnetic kink crystal in the chiral helimagnets under the transverse magnetic field, and
(c) concomitant phase modulation. In (b), we  depict a linear array of the spins along one chiral axis that is ferromagnetically coupled to the neighboring arrays. 
}\label{fig:CSL} 
\end{figure}

In the continuum limit, the Hamiltonian density corresponding to the lattice Hamiltonian (\ref{lattH}) is written as
\be
{\cal{H}}&=&
{1\over 2}
\left({\partial_x\theta}\right)^2
+{1\over 2}\sin^2\theta\left({\partial_x\varphi}\right)^2\non\\ 
&-&q_0\sin^2\theta\left({\partial_x\varphi}\right)-\beta\sin\theta\cos\varphi,\label{eqn1}
\en
where the energy is measured by $JS^2$, and $\beta=\tilde{H}/{{J}S}$. 
The semi-classical spin variable is represented as 
$
\bb{S}=
S(\cos\theta,
\sin\theta\cos\varphi,
\sin\theta\sin\varphi
)
$ by using the slowly varying polar angles $\theta(x)$ and $\varphi(x)$[see Fig.~\ref{fig:CSL}(a)].
The helical pitch for the zero field ($\beta=0$) is given by $q_0=\tan^{-1}(D/J)\approx D/J$. 
Under the transverse field, a regular array of the magnetic kink is formed.\cite{Dzyaloshinskii64}
The kink corresponds to the phase winding in the left-handed ($\Delta\varphi=+2\pi$) or right-handed($\Delta\varphi=-2\pi$) manners.
Since we assume the uniform mono-axial Dzyaloshinskii vector $\bm{D}$,
the kink with only positive (left-handed) or negative (right-handed) charge are energetically favored. The kink
 with the same charge repel each other, just like in the case of the Coulomb repulsion. So, the magnetic kink crystal (soliton lattice) is formed, 
as shown in Figs.~\ref{fig:CSL}(b) and (c).

The magnetic kink crystal phase  is described by the stationary soliton solution,
$\theta=\pi/2$ and 
$
\cos\left(\varphi_0(x)/ 2\right)={\rm{sn}}({m} x/\kappa),
$
where 
$
m=\sqrt{\beta}
$ 
corresponds to the first breather mass and 
\lq\lq$\rm{sn}$\rq\rq  denotes a Jacobian elliptic function.\cite{Dzyaloshinskii64}
The period of the kink crystal is given by
$
l_0={2\kappa K(\kappa)}/{\sqrt{\beta}}.
$
The elliptic modulus $\kappa$ ($0<\kappa<1$) is determined by
the energy minimization condition
$
{E(\kappa)/ \kappa}={\pi q_0/ 4m}.
$
Here, 
$K(\kappa)$ and
$E(\kappa)$ 
 denote the elliptic integrals of the first and second kind, respectively.



Now, we consider the fluctuations around the classical solution
and write
$
\theta(x)=\pi/2+u(x)$,
and
$
\varphi(x)=\varphi_0(x)+v(x).
$
When we consider only the tangential $\varphi$-mode,
our problem is reduced to the one first investigated by Sutherland.\cite{Sutherland73}
 The  $\varphi$-mode is fully studied  in the context of the chiral helimagnet.\cite{Izyumov-Laptev86,Aristov-Luther03}
In the present work, however, it is essential to take into account not only the 
$\varphi$-mode but the $\theta$-mode to argue the longitudinal magnetic current.
Expanding (\ref{eqn1}) up to $u^2$ and $v^2$, we have
$
H=\int dx({\cal{H}}_0+{\cal{H}}_u+{\cal{H}}_v+{\cal{H}}_{\rm{int}}),
$
where ${\cal{H}}_0$ gives the classical solution and
$
{\cal{H}}_u=
u{{\cal{L}}}_u u$,
${\cal{H}}_v=
v{{\cal{L}}}_v v,
$
where the differential operators are defined by
\be
{\cal{L}}_u&=& -{1\over 2}\partial_x^2-{1\over 2}(\partial_x\varphi_0)^2 
+q_0(\partial_x\varphi_0)+{1\over 2}\beta\cos\varphi_0,\\
{\cal{L}}_v&=&-{1\over 2}\partial_x^2+{1\over 2}\beta\cos\varphi_0.
\en
The lowest-order coupling between the $u$ and $v$ modes comes from
$
{\cal{H}}_{\rm{int}}=-u^2(\partial_x v)^2/2,
$
that is neglected here.
In the case of zero-field, $\beta=0$, 
 we have
$
{\cal{L}}_u= -\partial_x^2/2+q_0^2/2$, and
$
{\cal{L}}_v=-\partial_x^2/2.
$
Therefore we see that $u$-mode acquires the mass $q_0$ (scaled by $JS^2$), while
the $v$-mode becomes massless.
This situation naturally arises, because 
the $v$-mode is a Goldstone mode,
but the $u$-mode is not.
Even after switching the perpendicular field, the $u$-mode ($v$-mode) remains to be massive (massless).

\begin{figure}
\includegraphics[width=70mm]{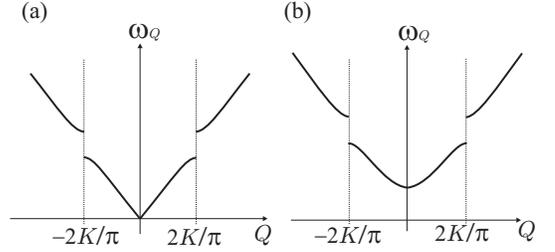}
\caption{ The energy dispersions of the eigen modes for 
(a) the tangential
$\varphi$-fluctuation ($\omega_Q=\sqrt{\rho}$) and (b) the longitudinal $\theta$-fluctuations  ($\omega_Q=\sqrt{\lambda}$). }
\label{fig:dispersion} 
\end{figure}

From now on, we argue that the massive $\theta$-fluctuations carry the magnetic current.
First, we perform the mode expansions,
$v(x,t)=\sum_n\eta _n(t)v_n(x)$ and
$
u(x,t)=\sum_n\xi _n(t)u_n(x)
$,
and seek the energy dispersions for the normal vibrational modes, satisfying
${\cal{L}}_u u_n(x)=\lambda_nu_n(x)$, and
${\cal{L}}_v v_n(x)=\rho_n v_n(x)$, respectively.
Introducing
$\tilde{x}=mx/\kappa$,
we have the Schr\"odinger-type equations,
\begin{eqnarray}
{d^2u_n(x)/ d\tilde{x}^2}&=&[2\kappa^2 \sn^2\tilde{x}\non\\
&&\!\!-{\kappa^2}(1+{2\lambda_n/  \beta})
-4+4 \kappa \tau ]u_n(x),\label{eveq:u}\\
{d^2v_n(x)/ d\tilde{x}^2}&=&
[2\kappa^2 \sn^2\tilde{x}-{\kappa^2}
(1+{2\rho_n/  \beta})
]v_n(x),\label{eveq:v}
\end{eqnarray}
with $\tau=q_0/m$.
In Eq.~(\ref{eveq:u}) we consider the case of weak field corresponding to small $\kappa$, leading to
$\dn x\approx 1$.
Now, the equations (\ref{eveq:u}) and (\ref{eveq:v})  reduce to
the Jacobi form of the Lam$\acute{\rm{e}}$ equation,\cite{WW}
$
{d^2{\Lambda}_{{\alpha}}(x)/ dx^2}=\left\{\ell(\ell+1)\kappa^2\sn^2 x+A\right\}{\Lambda}_{{\alpha}}(x),\label{lame}
$
with $\ell=1$.
It is known that the solution is parameterized by a single
continuous complex parameter $\alpha$
as,\cite{Sutherland73,Izyumov-Laptev86}
\begin{equation}
\ds
{\Lambda}_{{\alpha}}(x)=N
{
\vartheta_4\left(\pi (x-x_0) / 2K \right)
\over
\vartheta_4\left(\pi   x/2K \right)
}e^{-i{Q}x}
,\label{LamesolI}
\end{equation}
where $N$ is a normalizing factor and $\vartheta _i$ ($i=1,2,3,4$) denote the Theta functions\cite{WW} with
$Q$ being the Floquet index.\cite{Sutherland73,Izyumov-Laptev86,Aristov-Luther03} The energy dispersion is obtained by
determining $A=-\kappa^2(1+\tilde{A})$ as a function of the
Floquet index $Q$ which marks eigenstates instead of $n$. It is known\cite{Sutherland73} that the dispersion consists of two
(generally $\ell+1$) bands specified by the acoustic branch $\tilde{A}_1=\kappa
^{\prime 2}/\kappa ^2\,\sn^2\left( \alpha ,\kappa ^{\prime }\right) $, $Q_1=\pi
\alpha /2KK^{\prime }+Z\left( \alpha ,\kappa ^{\prime }\right) $, and the
optical branch $\tilde{A}_2=1/[\kappa ^2\sn^2\left( \alpha ,\kappa ^{\prime
}\right)] $, $Q_2=\pi \alpha /2KK^{\prime }+Z\left( \alpha ,\kappa ^{\prime
}\right) +\dn\left( \alpha ,\kappa ^{\prime }\right) \cn\left( \alpha ,\kappa
^{\prime }\right) /\sn\left( \alpha ,\kappa ^{\prime }\right) $, where $%
\alpha \in \left( -K^{\prime },K^{\prime }\right] $. 
Here, $K^{\prime }$ denotes the elliptic integral of the first kind with the complementary modulus
$\kappa'=\sqrt{1-\kappa^2}$ and $Z$ denotes the Zeta-function.\cite{WW}
The complex parameter $x_0$ in
Eq.(\ref{LamesolI}) are given by $i\alpha +K$ and $i\alpha $ for the acoustic and optical
branches, respectively.
We have the acoustic branch,
$
0\leq \tilde{A}_1 <{\kappa'^2/ \kappa^2}$,
for $
0\leq |{Q}_1| \leq {\pi/ 2K}
$, and the optical branch,
${1/ \kappa^2}\leq \tilde{A}_2 <\infty$,
for
${\pi/ 2K}\leq |{Q}_2|$.
The energy gap 
$
\Delta =1,
$
opens at $|Q|={\pi/ 2K}.$
We present the dispersions $\omega_Q$ in Fig.~\ref{fig:dispersion}, where the gapless acoustic $\sqrt{\mu_BHS}\tilde{A}_1^{1/2}$  and optical $\sqrt{\mu_BHS}\tilde{A}_2^{1/2}$ bands of $\varphi$-excitations are depicted together with the gapfull  acoustic $\sqrt{\mu_BHS} [\tilde{A}_1+4 \tau/\kappa -4/\kappa^2]^{1/2}$  and optical $\sqrt{\mu_BHS} [ \tilde{A}_2+4 \tau/\kappa -4/\kappa^2]^{1/2}$  bands of $\theta$-excitations.

Next we consider the collective dynamics of the kink crystal.
For this purpose, we carry out
the canonical formulation by using the collective coordinate method.\cite{Christ-Lee75}
We start with the corresponding Lagrangian,
\begin{eqnarray}
{{L}}=\int dx
\left[
S( 1-\cos \theta) \varphi _t-{\cal{V}}\left( \theta,\varphi \right)\right] , \label{L0}
\end{eqnarray}
where the Berry phase term is taken into account.
The mode expansion leads to the vibrational  term, ${\cal{V}}\left[ \theta ,\varphi \right]
=\sum_n\lambda _n\xi _n^2+\sum_n\rho _n\eta _n^2$.
Elevating the position of the kink center $X$ to a dynamical variable, we write the solution in the form 
\be
\left.
\begin{array}{c}
 \varphi =\varphi _0\left[ x-X(t)\right] +\sum_{n=1}^\infty \eta
_n(t)v_n\left( x-X(t)\right) , \\
 \theta =\pi /2+\sum_{n=1}^\infty \xi _n(t)u_n\left( x-X(t)\right).
\end{array}
\right\}
\en
Plugging these expressions into the Lagrangian (\ref{L0}), we have
$
L=-S\sum_n\dot{\eta}_n(t)K_{1n}+S\dot{X}\sum_nK_{2n}\xi
_n(t)
-S\sum_{n,m}K_{3nm}\xi _n(t)\dot{\eta}_m(t)
-\sum_n\lambda _n\xi_n^2-\sum_n\rho _n\eta _n^2,
$
with the coefficients 
$
K_{1n}=\int dxv_n\left( x\right)
$, 
$
K_{2n}=\int dx\left({\partial \varphi _0}/{\partial x}\right)
u_n\left( x\right)$, and
$
K_{3nm}=\int dxu_m\left( x\right) v_n\left( x\right) . 
$
This Lagrangian is {\it singular} because the
determinant of the matrix of second derivatives of the Lagrangian with
respect to velocities (Hessian) turns out to be zero.
Therefore we need to construct the Hamiltonian by using the Dirac's prescription for the constrained Hamiltonian systems.
The canonical momenta conjugated to
the coordinates $X$, $\xi _n$, and $\eta _n$, i.e.,
$
p_1={\partial L}/{\partial \dot{X}}=S\sum_nK_{2n}\xi _n, 
$
$
p_{2n}={\partial L}/{\partial \dot{\xi}_n}=0, 
$
$
p_{3n}={\partial L}/{\partial \dot{\eta}_n}=-SK_{1n}-S\sum_mK_{3mn}\xi
_m,
$
lead to
the extended Hamiltonian,
$
H^{*}=p_1\dot{X}+\sum_np_{2n}\dot{\xi}_n+\sum_np_{3n}\dot{\eta}_n-L
$
with a set of primary constraints,
\begin{eqnarray}
\left.
\begin{array}{c}
\Phi _1^{(1)}=p_1-S \sum_n K_{2n}\xi _n=0,  \\
\Phi _{2n}^{(1)}=p_{2n}=0, \\
\Phi _{3n}^{(1)}=p_{3n}+SK_{1n}+S\sum_mK_{3mn}\xi _m=0.
\end{array}
\right\}
\end{eqnarray}
Because of a lack of primary expressible velocities the Hamiltonian 
with the imposed constraints 
\[
H^{(1)}=\Phi _1^{(1)}\dot{X}+\sum_n\Phi _{2n}^{(1)}\dot{\xi}_n+\sum_n\Phi
_{3n}^{(1)}\dot{\eta}_n+H_{ph}, 
\]
coincides with $H^{*}$, where $H _{ph}=\sum_n\lambda _n\xi _n^2+\sum_n\rho _n\eta _n^2$.  
It governs the equations of motion of the constrained system, 
i.e. the constraints are hold at all times. This  leads to a set of  dynamical equations,
\be
\left.
\begin{array}{c}
\sum_nK_{2n}\dot{\xi}_n=0,  \label{me1}\\
-2\lambda _n\xi _n+\dot{X}SK_{2n}-S\sum_mK_{3nm}\dot{\eta}_m=0,  \label{me2}\\
-2\rho _n\eta _n+S\sum_mK_{3mn}\dot{\xi}_m=0,  \label{me3}
\end{array}
\right\} \label{c2}
\en
to give $\dot{\xi}_n=0$ and $\eta_n=0$.
Imposing the secondary constraints $\Phi _{n}^{(2)}=\eta_n=0$  to be constant in time,
we obtain  $\dot{\eta}_n=0$. Together with the second constraint in (\ref{c2})  this yields
$
\xi _n=({SK_{2n}}/{2\lambda _n})\dot{X},
$
and we reach the final form of the physical Hamiltonian,
$
H_{ph}=p_1^2/2M,
$
where $p_1=M\dot{X}$ involves the soliton mass 
\begin{equation}
M=S^2\sum_n\frac{K_{2n}^2}{2\lambda _n}.  \label{mass}
\end{equation}


Now we are ready to define the longitudinal spin current.
We start with the linear momentum carried by the kink crystal, 
\begin{equation}
P=S\int_0^{L_0}\left( 1-\cos \theta \right) \varphi _x\,dx,  \label{macroP}
\end{equation}
where $L_0$ is the system size.
By using $\theta =\pi /2+u$ and $\varphi =\varphi _0$, 
for a steady current, we obtain 
\be
P&\approx& S\left[ \varphi _0({L_0})-\varphi _0(0)\right] +S\int_0^{L_0}u(x)\frac{\partial
\varphi _0}{\partial x}\,dx\non\\
&=& 2\pi {\cal{Q}} S+S\sum\limits_n\xi _nK_{2n}.
\en
We here introduced  the topological charge, ${\cal{Q}}=[\varphi _0(L)-\varphi _0(0)]/2\pi$.
Using the result  $
\xi _n=({SK_{2n}}/{2\lambda _n})\dot{X}
$ and Eq.(\ref{mass}), we obtain an important formula, 
\be
P= 2\pi S {\cal{Q}} +M\dot{X},  \label{momentum}
\en
that plays an essential role in this paper.
The first term is associated with the equilibrium background momentum and 
the second one corresponds to the transport current carried by the $\theta $%
-fluctuations. Apparently, the transverse magnetic field increases
a period of the kink crystal lattice and diminishes the   topological charge 
${\cal{Q}}$ and therefore it affects only the background linear momentum. The physical
momentum related with a mass transport due to the excitations around the
kink crystal state is generated by the steady movement.

 The \lq\lq superfluid mass current\rq\rq \,is accompanied by 
the \lq\lq superfluid magnetic
current\rq\rq \,  transfered by the $\theta $-fluctuations. It is determined through the
definition of the magnetic density,\cite{Volovik07} 
$
{\cal{N}}=S( 1 - \cos\theta).
$
By using $\theta =\pi /2+u(x,t)$, we have
$
{\partial \cal{N}}/{\partial t}=S\sin \theta 
{\partial \theta }/{\partial t%
}\approx S{\partial u}/{\partial t},
$
with
$
u(x,t)=\sum_n\xi _n(t)u_n\left[ x-X(t)\right].
$
Therefore, 
for a steady current, we obtain  the continuity equation 
\be
\frac{\partial \cal{N}}{\partial t}=-\dot{X}^2\frac \partial {\partial x}\left(
\sum_n\frac{S^2K_{2n}}{2\lambda _n}u_n\right) =-\frac{\partial j^x}{\partial x%
},
\en
where we introduced the magnon time-even current carried by the $\theta $-fluctuations
\be
j^x=S^2\dot{X}^2\sum_n\frac{K_{2n}}{2\lambda _n}u_n.\label{lc}
\en
Here we used the fact $\dot{\xi}_n=0$ because of the constraint.
The time evenness is manifested by appearance of not $\dot{X}$ but $\dot{X}^2$.
The important point to note is that {\it the only massive $\theta$-mode can carry the 
longitudinal magnon current as a  manifestation of ordering in non-equilibrium state, i.e.,
dynamical off-diagonal long range order.
}

\begin{figure}[h]
\includegraphics[width=75mm]{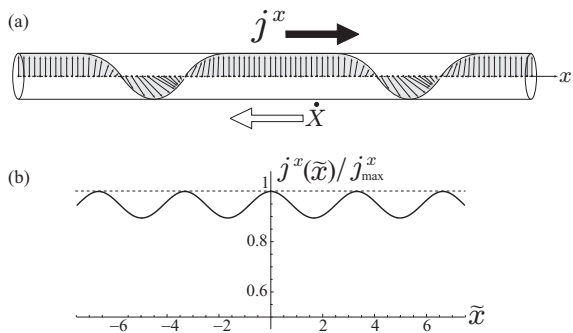}
\caption{(a) A schematic view of an instant distribution of spins in the current-carrying state. This picture 
corresponds to the case of intermediate field strength. 
(b) A snapshot of the position dependence of the current density $j^x(\tilde{x})$ in the weak 
field limit, exactly treated in this paper.
$j^x(\tilde{x})$ is scaled by its maximum $j_{\rm{max}}^x=j^x(0)$.
}
\label{Volovik}
\end{figure}

 The final stage is to explicitly compute (\ref{lc}). 
We can exactly prove that the optical branch does not contribute 
to the magnetic current because of the orthogonality (the proof is detailed in a later paper). 
After a lengthy but straightforward manipulation, the contribution of the acoustic branch is obtained
as a function of $\tilde{x}=mx/\kappa$,
\begin{eqnarray}
j^x(\tilde{x})
&=&
\frac{8E^2(\kappa ){\dot{X}^2}}{J\pi ^2q_0^{2}\left( 4E(\kappa )/\pi
-1\right) }\;\dn\left(\tilde{x}\right) 
\approx
\frac{2{\dot{X}^2}}{Jq_0^{2}}\;\dn\left(\tilde{x}\right) ,\non\\
\label{AcSpCur}
\end{eqnarray}
for the weak field case corresponding to small $\kappa$ leading to $E(\kappa )\approx \pi /2$.
On the other hand, the background spin current\cite{Heurich03} is shown to become   
$j_{\rm{bg}}(x)\approx\partial \varphi_0(x)/\partial x-q_0
\propto
\dn(\tilde{x})-{2} E(\kappa)/\pi$. We stress that the physical meaning of $j_{\rm{bg}}$ is completely different from the current described by Eq.~(\ref{AcSpCur}) in {\it non-equilibrium} state. \cite{Xiao}
%
%
%
 We present a schematic view of an instant distribution of spins in the current-carrying state in Fig.~\ref
{Volovik}(a). 
In Fig.~\ref{Volovik}(b), we  present a snapshot of the position dependence of the current density 
$j^x(\tilde{x})=j^x_{\rm ac}(\tilde{x})$ in the weak field limit, given by Eq.~(\ref{AcSpCur}).

In realizing the bulk magnetic current proposed here,
a single crystal  of chiral magnets serves as spintronics device.
The mechanism involves no spin-orbit coupling and 
the effect is not hindered by dephasing.
Finally, we propose possible experimental methods to trigger off the spin current  considered here. 
{\it Spin torque mechanism}: the spin-polarized electric current can exert torque to ferromagnetic moments through 
direct transfer of spin angular momentum.\cite{Slonczewski96} This effect, related with Aharonov-Stern effect \cite
{Aharonov-Stern} for a classical motion of magnetic moment in an inhomogeneous magnetic field, is eligible to excite 
the sliding motion of the kink crystal by injecting the spin-polarized current (polarized electron beam) in the 
direction either perpendicular or  oblique to the chiral axis. The spin current transported by the soliton lattice may 
amplify the spin current of the injected carriers.
{\it XMCD}: to detect the longitudinal magnetic currents accompanied by the dynamical ODLRO, x-ray magnetic circular 
dichroism (XMCD) may be used.  Photon angular momentum may be aligned either parallel or anti-parallel to the 
direction of the  longitudinal net magnetization. 
{\it Ultrasound}: further possibility to control and  detect the spin current is using a coupling between spins and 
chiral torsion.\cite{Fedorov} Ultrasound with the wavelength being adjusted to the period of the kink crystal may 
excite the periodic chiral torsion and  resonantly supply  the kinetic energy to the kink crystal. Consequently, the 
ultrasound attenuation may occur.\cite{Hu}
{\it TOF technique}: the most  direct way of detecting the traveling magnon density  may be winding a sample by 
a pick-up coil and performing the time-of-flight (TOF) experiment. Then, the coil should detect
a periodic signal induced by  the magnetic current.

\begin{acknowledgments}
We acknowledge helpful discussions with
Yu.~A.~Izyumov, K.~Inoue, I.~Fomin and M.~Sigrist. 
J.~K. acknowledges Grant-in-Aid for Scientific Research (A)(No.~18205023)
and (C) (No.~19540371) from the Ministry of
Education, Culture, Sports, Science and Technology, Japan.
\end{acknowledgments}


\end{document}

\bibitem{Rubinstein70}
J. Rubinstein, J. Math. Phys. \textbf{11}, 258
(1970).

\bibitem{Dirac} P.A.M. Dirac, Lectures on Quantum Mechanics, (Yeshiva, New York, 1964).

\bibitem{Canali} C. Canali, C. Jacoboni, F. Nava, G. Ottaviani, and A. Alberigi-Quaranta,  Phys. Rev. B \textbf{12}, 2265 (1975).

\bibitem{Izyumov1985}
Y.~A.~Izyumov:
Sov. Phys. Usp. \textbf{42} (1985) 845.

\bibitem{Izyumov1991}
Yu.~A.~Izyumov:
Physica \textbf{174B} (1991) 9.

{\it Wiedemann effect}: magneto-mechanical devices using Wiedemann effect have been recently proposed \cite{Wiedemann}, where the spin current is detected by mechanical torque.  Correspondingly, the inverse Wiedemann effect (IWE) may also be used to detect spin current. The longitudinal magnetization in a wire of chiral magnet subject to weak torsional stress will cause circular distribution of the magnetic field[Hernando].

\bibitem{Wiedemann} P. Mohanty, G. Zolfagharkhani, S. Kettemann, and P. Fulde, Phys. Rev. B \textbf{70}, 195301 (2004); A.G. Mal'shukov, C.S. Tang, C.S. Chu, and K.A. Chao, Phys. Rev. Lett. \textbf{95}, 107203 (2005).

\bibitem{Liu73}
L.~L.~Liu,
Phys. Rev. Lett. \textbf{31}, 459 (1973).

the domain wall motion is accompanied by the emergence of the longitudinal magnetic moment and

\bibitem{Volovik07} 
This point was clearly recognized in the context of the spin supercurrent in the  superfluid $^3$He.
See, I.A. Fomin, Physica
\textbf{B169}, 153 (1991);
G.~E.~Volovik,  arXiv:cond-mat/0701180.

In our proposal, not heterostructure or multilayers but a single crystal  of chiral magnets serves as spintronics device, the crystal symmetry ensures itself  the chiral structure.  The mechanism involves no spin-orbit coupling that causes the spin dephasing.